\documentclass[11pt,nofootinbib,showkeys,preprintnumbers]{revtex4}
\usepackage{epsfig}
\usepackage[dvipdfm]{hyperref}
\usepackage{amssymb}
\usepackage{amsbsy}

\def\tit{\subsection}

\def\e{{\,\rm e}}
\def\be{\begin{equation}}
\def\ee{\end{equation}}
\def\bea{\begin{eqnarray}}
\def\eea{\end{eqnarray}}
\def\LA{\left\langle}
\def\RA{\right\rangle}
\newcommand{\rf}[1]{(\ref{#1})}
\newcommand{\eq}[1]{Eq.~(\ref{#1})}

\renewcommand{\!}{\negthinspace}

\def\i{{\rm i}}

\def\K{{K}}
\def\N{{N}}

\def\t{{s}}
\def\d{{\rm d}}
\def\D{{\cal D}}
\def\Tau{{\cal T}}
\def\x{x}
\def\r{r}

\newcommand{\pinti}{\int_{-\infty}^{+\infty}\hspace{-2.62em}\not
\hspace{2.2em}}

\newcommand{\ie}{{{\it i.e.\/ }}}
\def\la{\lesssim}
\def\ga{\gtrsim}

\newcommand{\non}{\nonumber \\*}

\begin{document}

\title{\large\sf{\bf Scattering amplitudes of QCD string}}
\author{{Yuri Makeenko}} 
\address{Institute of Theoretical and Experimental Physics, 
B. Cheremushkinskaya 25, 117218 Moscow, \\E-mail: makeenko@itep.ru}

\begin{abstract}

{\small
I review the derivation of large-$N$ QCD 
meson scattering amplitudes in the Regge regime, where the effective 
theory of long strings applies in $d=4$. 
A special attention is payed to the reparametrization path integral
which plays a crucial role in the consistency of off-shell amplitudes.
I show how the linear Reggeon trajectory is obtained for QCD string in 
the mean-field approximation, which
turns out to be exact for the Nambu--Goto string, and discuss
the interrelation with perturbative QCD.

Based on the talk given at the Workshop
``Scattering Amplitudes: from QCD to maximally supersymmetric 
Yang-Mills theory and back'', Trento 16--20 July 2012}.

\end{abstract}


\maketitle

\section{Introduction}

QCD string is made from fluxes of the gluon field 
and makes sense for the distances larger than the confinement scale. 
There are two cases where QCD string is described by an effective theory 
of long strings:
the static potential (reviewed in Y.M. (2012) \cite{Mak12})
and meson scattering amplitudes in the Regge regime
described in this talk. I review the results of the References 
\cite{MO08,MO09,BM09,MO10a,MO10b,Mak11a,Mak11b,MO12}, where
the scattering amplitudes of QCD string were obtained in the Regge regime,
using the worldline formalism.

For the consistency of the (off-shell) amplitudes, 
a very important role is played by the 
reparametrization path integral (a synonym of the path integral over 
boundary metrics or the path integral over boundary values of the 
Liouville field). This issue is reviewed in the first part of
this talk.

The second and third parts are devoted to 
 scattering amplitudes of fundamental string and QCD string
in the {Regge} regime. I consider  (polygonal)
 momentum Wilson loops, semiclassical fluctuations about
the associated minimal surface and the mean-field approach
to the Reggeon trajectory in $d=4$. I also discuss
off-shell scattering amplitudes of fundamental string and
the relative contribution of perturbative QCD to QCD string.

\section{Reparametrization path integral}

\tit{Path integral over reparametrizations}

It is commonly believed that the
Wilson loop of large size in large-$N$ QCD 
equals the string disk amplitude. As was emphasized
by {Polyakov (1997)} \cite{Pol97}, it is important to path
 {integrate} over {reparametrizations} of the boundary:
\be
W[x(\cdot)]= \int {\cal D}_{\rm diff} t(s)\e^{-\K S[x(t)]}\,,
\label{W[x]}
\ee
\ie over functions $t(s)$ with $\dot t(s)\equiv \d t(s)/\d s\geq 0$
({here $\K=1/2\pi{\alpha'}$ is  the string tension}).

The necessity for reparametrizations of the boundary curve within
the Polyakov string formulation was pointed out 
long ago by Polyakov (1981) \cite{Pol81}
and Alvarez (1983) \cite{Alv83}. The path integral over reparametrizations
first appeared for an off-shell propagator in
{Cohen, Moore, Nelson, Polchinski (1986)} \cite{CMNP86}.

The boundary action $S[x(t)]$ in \eq{W[x]} reads explicitly
\be
S[x(t)]=\frac{1}{4\pi}
\int_{-\infty}^{+\infty}\frac{\d s_1 \d s_2}{(s_1-s_2)^2} \,
{\left[x(t(s_1))-x(t(s_2))\right]^2}
\label{SS}
\ee
for the upper half-plane parametrization of the string worldsheet, where
the boundary is parametrized by the real axis.
It is known from {Douglas (1931)} \cite{Dou31} {algorithm} 
of solving the {Plateau problem}
(finding the minimal surface), which prescribes
to {minimize} the {boundary functional} \rf{SS}
with respect to {reparametrizations} $t(s)$. 
The boundary curve in \eq{SS} is fixed,
so the boundary action is a functional of $t(s)$.

A few comments are in order:
\begin{itemize}
\vspace*{-7pt}
\addtolength{\itemsep}{-7pt}

\item The representation \rf{W[x]} can be derived for the Polyakov string in 
critical dimension $d=26$ 
by doing the Gaussian path integral over string worldsheets with fixed
boundary.
The Liouville field $\varphi$,
 which enters through the conformal factor of
the 2D metric tensor $g_{ab}=\e^{\varphi} \delta _{ab}$, decouples then
in the bulk,
while its boundary value is related to the reparametrizations as
\be
\frac{\d t(s)}{\d s}=\e^{\varphi (s,0)/2}.
\ee
Thus, the path integral over reparametrizations is the same as
that over boundary metrics.

\item For the function  $t(s)=t_*(s)$, minimizing Douglas' integral \rf{SS},
the embedding-space coordinates obey the
conformal gauge (\ie the Virasoro constraints), 
so the quadratic stringy action  coincides
with the Nambu--Goto action. This is why the boundary action \rf{SS} 
reproduces the minimal area.

\item The area law  for asymptotically {large} loops
can be obtained as 
a {saddle point} in the {reparametrization} path integral \rf{W[x]}
at $t(s)=t_*(s)$. It can be shown that the 
{zig-zag} (or {backtracking) symmetry} holds,
as it should for the minimal area, owing to an explicit form of $t_*(s)$.

\item The coordinates $x^\mu(t)$ enter 
the boundary action~\rf{SS} quadratically, which makes it easy to further
integrate over the boundary curves. The nonlinearities of the problem
then reside in the reparametrization path integral.

\item In  $d<26$  the ansatz~\rf{W[x]} has to be
modified by incorporating the path integral over
bulk fields, as is given by the effective string theory
of Polchinski, Strominger (1991) \cite{PS91}
reviewed in Y.M. (2012) \cite{Mak12}. The reparametrization path integral
remains crucial for the consistency.

\end{itemize}

Integrating by parts, Douglas' integral can be rewritten as
\be
S[x(t(s))]=\frac 12
\int_{-\infty}^{+\infty}{\d t_1 \d t_2} \,
{\dot x(t_1)\cdot \dot x(t_2)} \, G\left( s(t_1)-s(t_2)  \right),\quad
 G\left( s_1-s_2\right)=-\frac1\pi \ln |s_1-s_2|
\label{SSs}
\ee
and the reparametrization path integral goes over the inverse functions
$s(t)$.

\tit{Nontrivial example: ellipse}


A simple nontrivial example, showing the need of the boundary reparametrization,
is an elliptic boundary curve. Using the
unit-disk parametrization $z=r\e^{\i \phi}$, we write it as
\be
x^1=a\cos \theta(\phi), \qquad x^2=b\sin \theta(\phi) ,
\ee
where $a$ and $b$ are the major and minor radii of the ellipse
and $\theta(\phi)$ with $\dot\theta(\phi)\geq0$ reparametrizes the boundary.

Suppose $\theta_*(\phi)=\phi$, then Douglas' integral equals
\be
S[x(\theta)]=\pi\frac{a^2+b^2}2 \qquad \hbox{rather than}\qquad  \pi a b.
\ee
The equality is only for a circle $a=b$, when the 
unit-disk coordinates are conformal (or {isothermal)}.

The minimization of 
Douglas' integral \rf{SS} for an ellipse gives 
for $\theta(\phi)$ the incomplete elliptic integral:
\be
\dot\theta_*(\phi)= 
\frac{\pi}{2K(\nu)}\frac{1}{\sqrt{(1-\nu)^2 +4\nu\sin^2 \phi}}\,,
\qquad 
\frac{\pi K\left(\sqrt{1-\nu^2}\right)}{2K\left(\nu\right)}=
\log \frac{a+b}{a-b}\,,  
\ee
where $ {K(\nu)} $ 
is the complete {elliptic integral} of the first kind. 
These formulas can be obtained by conformal mapping of a unit disk 
onto the interior of an ellipse. 

We explicitly see from this simple example how the reparametrization of the 
boundary is needed to recover the minimal area.

\tit{Large loops and minimal area}

The area law [= the exponential of (minus) the minimal area] is thus recovered 
from the reparametrization path integral \rf{W[x]} in the saddle-point
approximation.

{Gaussian fluctuations} about the saddle-point $t_*(s)$
result in a {preexponential factor}:
\be
W\left[x(\cdot)\right] \stackrel{{\rm large~loops}}=
F\left[\sqrt{\K}x(\cdot)\right]
\e^{-\K S_{\rm min}\left[x(\cdot)\right]}
\left[1+ {\cal O}\left((\K S_{\rm min})^{-1}\right)\right],
\label{circle1}
\ee
which is  {contour dependent}.
Its contribution {for large loops} is much 
less than that of the minimal area.

This preexponential factor shows up, however, in more subtle effects 
(such as the {L\"uscher term}), 
coming from fluctuations  about $t_*(s)$: 
\be
t(s)=t_*(s)+\frac{\beta(s)}{\sqrt{\K S_{\rm min}}}\,.
\ee 
For a $R\times T$ {rectangle} with $T R \gg 1/\K$ and $T\gg R$ 
it is possible to path integrate over $\beta(s)$, as is done in
Y.M., Olesen (2010) \cite{MO10b}, to obtain
\be
F\left[{{\rm rectangle}}\right] \propto \e^{\pi T/R} 
\qquad \hbox{for }T\gg R
\ee
reproducing the {L\"uscher term} for {bosonic string} in $d=26$.
To this order we can restrict ourselves with the quadratic approximation
in $\beta(s)$, so it is not essential what is the actual measure 
in \eq{W[x]}.

This demonstrates that the reparametrization path integral \rf{W[x]}
knows about bulk fluctuations.

\tit{Discretization of the measure}

To construct the measure on $Diff(\mathbb{R})$ in \eq{W[x]}, we split
the interval $[s_0,s_f]$ into $N$ pieces and define~\cite{MO09} 
\be
\int\limits_{\t(\tau_0)=\t_0 \atop \t(\tau_f)=\t_f} \D_{\rm diff}\t(\tau)\cdots
= \lim_{N\to\infty}\int_{\t_0}^{\t_f}  
\prod_{j=1}^{N-1}\int_{\t_0}^{\t_{j+1}} \d \t_j\, 
\frac{1}{(\t_{j+1}-\t_{j})}  
\frac{1}{(\t_{1}-\t_{0})}  \cdots \,,
\label{themea}
\ee
where the integration goes over $(N-1)$ {subordinated} values
$s_1$, \ldots,  $s_{N-1}$:
$\t_0 \leq \cdots \leq \t_{i-1} \leq \t_i \leq \cdots \leq \t_N=\t_f$.
Thus defined measure is covariant under {reparametrizations} 
\be
s\rightarrow t(s)\,,\quad t(s_0)=s_0\,,
\quad t(s_f)=s_f\,, \quad \frac {\d t}{\d s}\geq 0\,.
\ee

Discretizing  $s'=\exp[-\varphi]$ that relates {reparametrizations}
 to the {boundary value} of the {Liouville field} $\varphi$
by $s_i-s_{i-1}=\exp[-\varphi_i]$, we write the measure \rf{themea} as  
\be
\int_{s_0}^{s_f}\D_{\rm diff} s \cdots =
\lim_{N \to \infty}
\prod_{i=1}^{N} \int_{-\infty}^{+\infty} {\d \varphi_i}\,
\delta^{(1)}\Big( s_f-s_0-\sum_{j=1}^{N} \e^{-\varphi_j} \Big)\cdots 
\ee
with {the only} restriction on $\varphi_i$'s given by the delta-function.

The integral over $s_i$ in \eq{themea} is (logarithmically) divergent 
and can be nicely regularized by changing
\be
\frac{1}{(s_{i}-s_{i-1})} \longrightarrow 
\frac{1}{\Gamma(\delta_i)(s_{i}-s_{i-1})^{1-\delta_i}} \qquad \hbox{all}~
\delta_i=\delta\,.
\ee

The main integral for the integration at the {intermediate} point $\t_i$ is
\be
\int_{\t_{i-1}}^{\t_{i+1}}\d \t_i\,  
\frac{\Gamma^{-1}(\delta_i)\Gamma^{-1}(\delta_{i+1})}
{(\t_{i+1}-\t_i)^{1-\delta_{i+1}}(\t_{i}-\t_{i-1})^{1-\delta_i}} 
=\frac{\Gamma^{-1}(\delta_i+\delta_{i+1})}
{(\t_{i+1}-\t_{i-1})^{1-\delta_i-\delta_{i+1}}}\,.
\ee
This is an analogue of the well-known convolution formula
\be
\int_{-\infty}^{+\infty} \frac{\d \t_i }{\sqrt{2\pi}}
\frac{\e^{-(\t_f-\t_i)^2/2\nu_1}}{\sqrt{\nu_1}}
\frac{\e^{-(\t_i-\t_0)^2/2\nu_2}}{\sqrt{\nu_2}}
=\frac{\e^{-(\t_f-\t_0)^2/2(\nu_1+\nu_2)}}
{\sqrt{(\nu_1+\nu_2)}}
\ee
used for calculations of Feynman's path integral with the usual 
{Wiener} measure. 

The {functional limit} is reached when  $N\to\infty$ with $N\delta\to0$: 
\be
\int_{s_0}^{s_N=s_f}\D_{\rm diff}^{(N)} s = \frac{1}{\Gamma(N \delta)}
\frac{1}{\left(s_{N}- s_{0} \right)^{1-N\delta}}
 \stackrel{N\delta\to0}
\longrightarrow N \delta\frac{1}{\left(s_{f}- s_{0} \right)}\,,
\ee
reproducing the {projective-covariant} result. This is an analogue
of the free propagator.

\tit{Reparametrizations as L\'evy stochastic process}


What trajectories are typical in the path integral over reparametrizations?

To answer this question, 
Buividivich, Y.M. (2009)~\cite{BM09} consider a
subordinated {stochastic process} (called the {gamma-subordinator}) 
with the probability density function
\be
P\left(\Delta s_i \right) =\frac{1}{\Gamma(\delta)
\left(\Delta s_i \right)^{1-\delta}}
\label{PDF}
\ee
with $\delta>0$ being a {time step}. Then
\be
\d s_f \int_{s_0}^{s_f}\D_{\rm diff}^{(N)} s 
\ee
has the meaning of a propagator from $s_0$ to $ \left[s_f,s_f+\d s_f \right]$
during the {time} $\tau=N\delta$. 

Introducing the scaling variable
\be
z=\tau \ln \frac1{(s_f-s_0)}\,,
\ee 
which is analogous to $(s_f-s_0)^2/\tau$ for the Gaussian random walks,
we write
\be 
\frac{ \tau \d s_f}{\left(s_{f}- s_{0} \right)^{1-\tau}}
=\d z \e^{-z} \,.
\ee
Therefore, a scaling occurs with
\be
(s_f-s_0)\sim \e^{-1/\tau} 
\ee
and we conclude that the Hausdorff dimension $d_H=0$ in this case.
This
supersedes $(s_f-s_0)^2\sim\tau$ for the Gaussian stochastic process of
the {Brownian motion} (whose $d_H=2$).

\begin{figure}
\includegraphics[width=7.3cm]{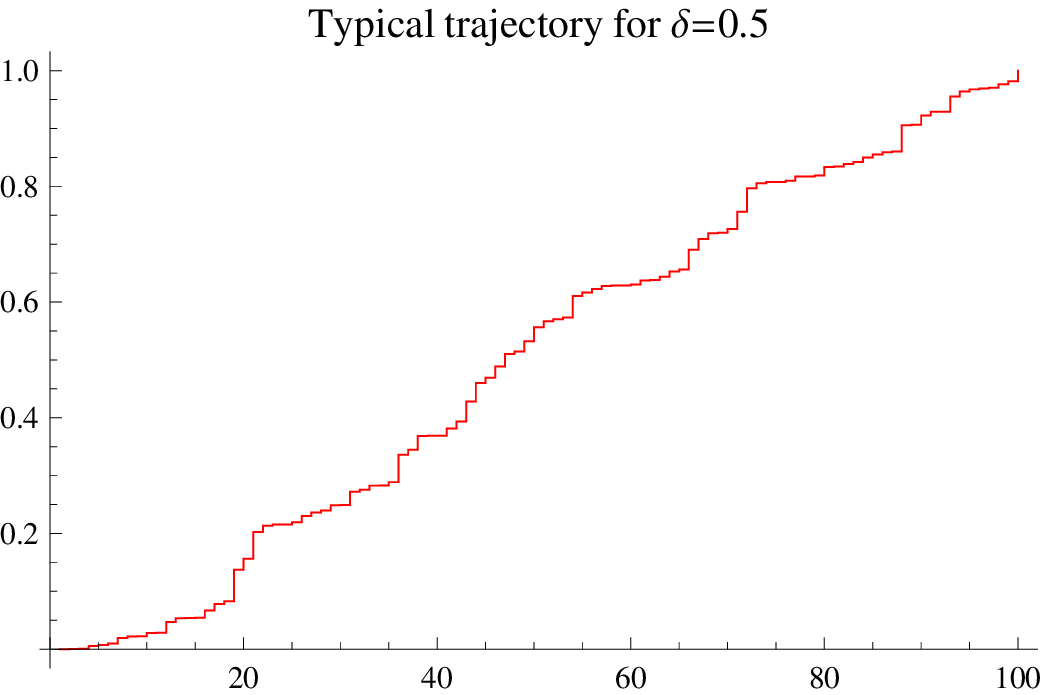}
\includegraphics[width=7.3cm]{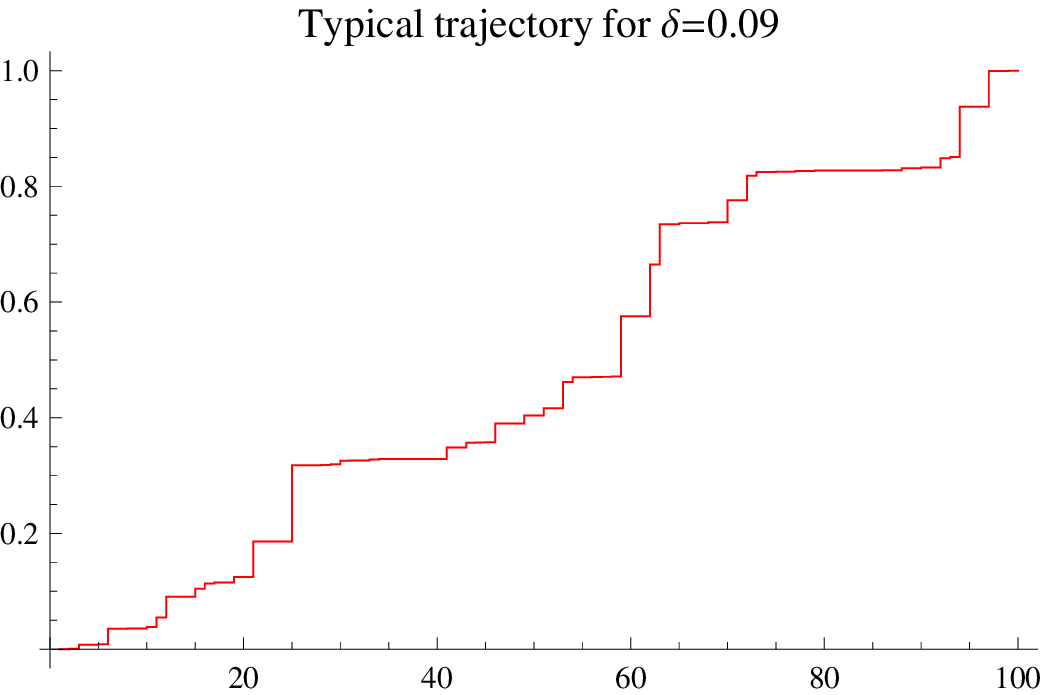}
\caption{Typical trajectories for the {gamma-subordinator} 
obtained by the Metropolis--Hastings algorithm. 
L\'evy's flights are seen in the right figure.}
\label{fi:1}
\end{figure}

Typical trajectories for the {gamma-subordinator} can be 
obtained by the {Metropolis--Hastings} algorithm in spite of
several subtleties, such as the central limit theorem and/or the
law of large numbers are not applicable for the probability
density \rf{PDF} which has an infinite dispersion. 
The results are depicted
in Fig.~\ref{fi:1} for $\delta=0.5$ and $\delta=0.09$.
{L\'evy's flights} are seen in the right figure. 
Their origin is that $P(\Delta s_i)$ is very large at small $\Delta s_i$
${\Longrightarrow}$ most of $\Delta s_i$'s are {small}. 
Then some of $\Delta s_i$ has to be
{large} to satisfy the {boundary condition}.

\tit{Hausdorff dimension of sample trajectories}

The {Hausdorff dimension} of the gamma-subordinator is expected to 
{decrease} from 1 to 0 with decreasing $\delta$,
as discussed by Horowitz (1968) \cite{Hor68}.

The {Hausdorff dimension} of the {discretized} process is determined
by its (L\'evy--Khintchine) {characteristic function}  
as
\be
d_H= \lim_{q\to\infty} 
\frac{\ln\left(-N \ln\LA \e^{-q\Delta s_{i}}\RA \right)}{\ln q}.
\label{dH}
\ee
This definition is equivalent to a more familiar one based
on the covering by balls.

\begin{figure}
\includegraphics[width=7.2cm]{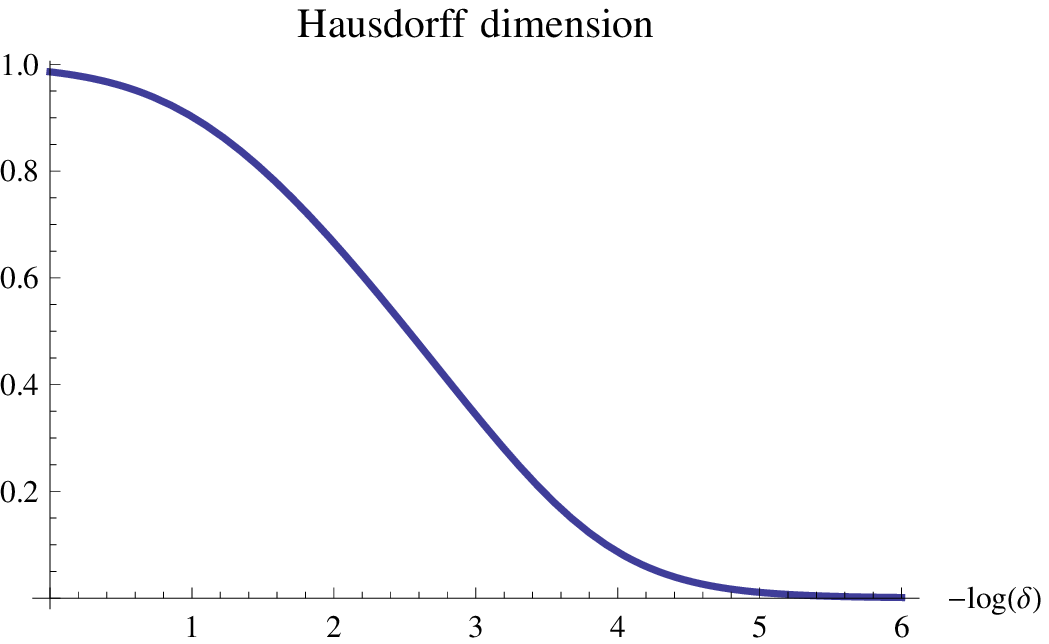} \hspace*{2mm}
\includegraphics[width=7.2cm]{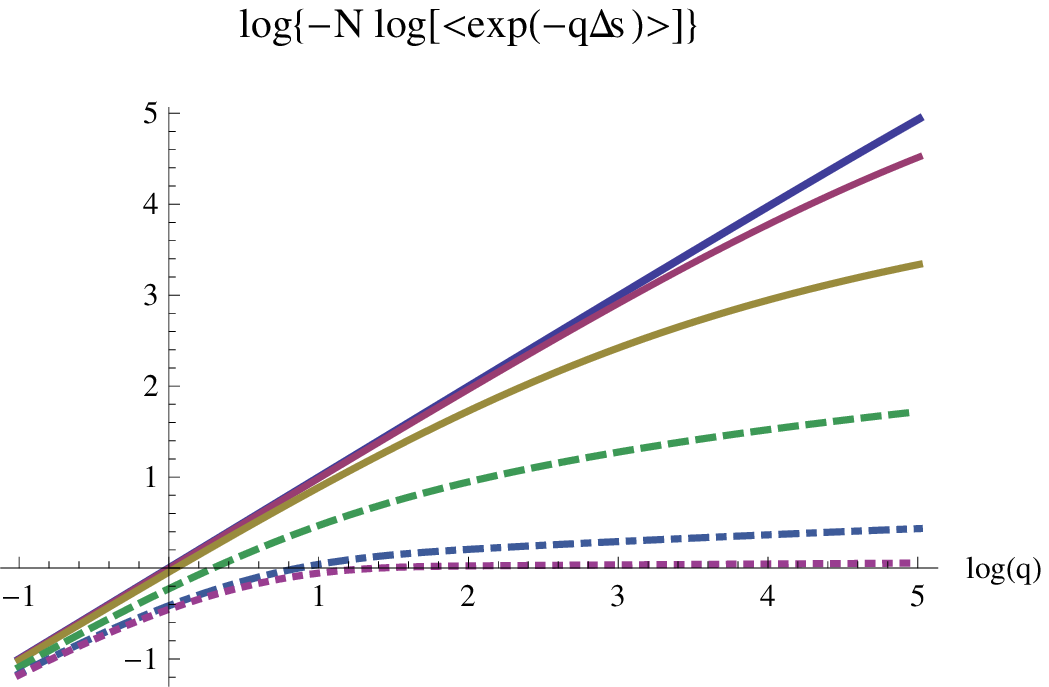}
\caption{Hausdorff dimension versus  $\ln(1/\delta)$ (left) extracted from the 
slope of the lines in the right figure for
$\delta=1,$  $10^{-1}$, $10^{-2}$, $10^{-3}$, $10^{-4}$, $10^{-5}$
from the top to the bottom.}
\label{fi:2}
\end{figure}

For the probability density~\rf{PDF} Buividivich, Y.M. (2009)~\cite{BM09} found
\be
\LA \e^{-q\Delta s_{i}}\RA ={}_1F_1(\delta, \delta N;-q),
\ee
where ${}_1F_1$ is the confluent hypergeometric function.
Substituting in \eq{dH}, we obtain the Hausdorff dimension
plotted  versus  $\ln(1/\delta)$ in Fig.~\ref{fi:2} (left).
The values of $d_H$ are extracted from the 
slope of the lines in Fig.~\ref{fi:2} (right).
The Hausdorff dimension decreases from 1 
for $\delta\ga 1$ to 0 for $\delta N\to0$, thus 
confirming the consideration of Horowitz (1968) \cite{Hor68}.

\tit{Ambiguities of the measure}

The discretization of the measure displayed on the right-hand side of 
\eq{themea} is not unique. A
more symmetric discretization of the measure reads
\be
\D_{\rm diff} s = \prod_{i=1}^{N-1} \d s_i 
\frac{(s_{i+1}-s_{i-1})}{(s_{i+1}-s_{i})(s_{i}-s_{i-1})}
\qquad\qquad \hbox{{Lovelace choice}}\,.
\label{Lovem}
\ee
Every multiplier is now
{invariant} under the $PSL(2;\mathbb{R})$ 
{projective transformations}
\be
\t\Rightarrow \frac{a \t+b}{c \t+d}\qquad \hbox{with}\quad  a d-b c=1 
\ee
which form a subgroup of the {reparametrizations}.
For the measure \rf{themea} only the product over $j$ was invariant
under  $PSL(2;\mathbb{R})$.

The measure \rf{Lovem} was considered in some detail in 
Y.M., Olesen (2010) \cite{MO10b} and
results in consistent off-shell (Lovelace) string amplitudes
of the intercept $\alpha(0)=(d-2)/24$ described below in Subsect.~\ref{ss:l}.

It is worth noting that
the results do not change if next-to-neighbor 
points are involved in the discretization, \ie
\be
(s_{i+1}-s_i)\Longrightarrow (s_{i+n}-s_i)/n\,.
\ee
This supports the expectation that 
a continuum limit exists in spite of the discontinuities
of the trajectories. However, the measures \rf{themea} and 
\rf{Lovem} apparently belong to
different {universality classes} {(which differ by the value of $\alpha(0)$)}.

\section{Scattering Amplitudes as Momentum Loops}

\tit{Momentum loops}

As was first pointed out by {Migdal (1986)} \cite{Mig86},
{scattering amplitudes} are given by a
{reparametrization-invariant} {functional Fourier transformation} 
\be
A[ p(\cdot)]= \int \D  x\, \e^{\i \int  p \cdot\d x }\; {J [x(\cdot)]}\,
W [x(\cdot)]\qquad \hbox{with process-dependent}\quad {J [x(\cdot)]}
\label{MigFF}
\ee
of the Wilson loop (to be identified with the {string disk amplitude})
for {a piecewise constant}
{momentum loop} 
\begin{figure}
\includegraphics[width=6cm]{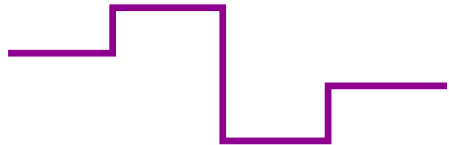} 
\caption{Piecewise constant momentum loop.}
\label{fi:pc}
\end{figure} 
\be
p^\mu(t)=P^\mu_i \qquad \hbox{for}~ t_i<t<t_{i\!+\!1} \,,
\label{stepW}
\ee
which is schematically depicted in Fig.~\ref{fi:pc}.

Differentiating the step function, we obtain
\be
\dot { p}^\mu(t) =- \sum_i   p_i^\mu \delta(t-t_i)
\ee
{with} $ p_i^\mu \equiv  P_{i\!-\!1}^\mu -  P_i^\mu$
representing $M$ momenta of (all incoming) particles.
Then {momentum conservation} is automatic while an (infinite) volume 
$V$ is produced, say, by integration over $\x_0=\x_M$. 

The {Fourier transformation} of the string {vertex operators} can be
 reproduced as follows. For scalars we simply  integrate by parts   
\be 
 \int \d t \,p(t) \cdot \dot x(t) =-
 \int \d t \, \dot p(t) \cdot x(t) =\sum_i  p_i \cdot \x_i \,.
\ee 
For vectors we first apply the variational derivative 
$\delta/\delta p^\mu(t)$ for an
arbitrary $p^\mu(t)$, which inserts $\i \dot x^\mu(t)$, and then set
$p^\mu(t)$ to be piecewise constant. It is similar for higher spins.


\tit{Momentum disk amplitude}

After the {Gaussian} path integration over $\D x^\mu(t)$
(which produces an {$s$-independent} determinant) we get the amplitude
\be
A[p(\cdot)]=  
\int \D_{\rm diff} s(t) \exp{\Big({{\alpha^\prime} }
\int\nolimits_{-\infty}^{+\infty} {\d t_1 \pinti\d t_2} \, 
\dot p(t_1) \cdot \dot p(t_2)\,
\ln |s(t_1)-s(t_2)|   \Big)}, 
\label{A[p]}
\ee
which looks like the {disk amplitude} \rf{W[x]} (the {Wilson loop})
for the boundary curve 
\be
 x^\mu(t)=\frac 1\K p^\mu(t).
\label{xp}
\ee
This can be seen by comparing the exponent in \eq{A[p]} with \rf{SSs}.

\begin{figure}
\centering{
\includegraphics[width=6cm]{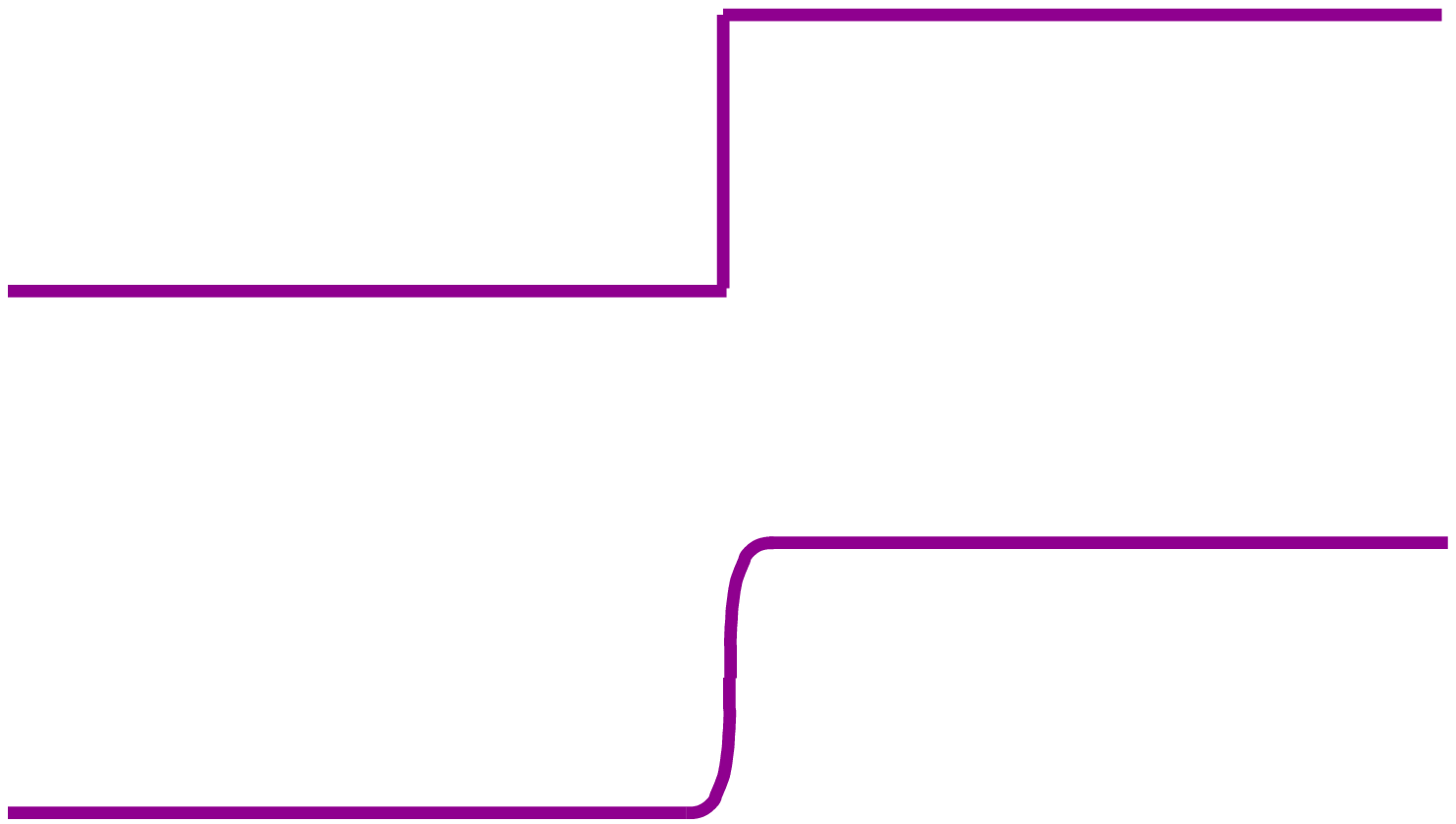}
\hspace*{16mm}
\includegraphics[width=3.8cm]{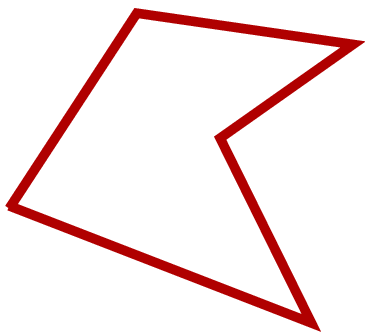}  }
\caption{Stepwise momentum loop (top left)
smeared with the width $\varepsilon_i$
(bottom left) and its polygonal image in embedding space (right).}
\label{fi:sme}
\end{figure}

Actually, the discontinuities of  the stepwise 
momentum loop are always smeared by a regularization
which involves the boundary {Liouville field}  $\varphi_i=\varphi(s_i,0)$ for 
the {covariance} as will be momentarily discussed:
\be
p^\mu(t)=
\frac 1{\pi} \sum_i p^\mu_i  \arctan \frac{(t-t_i)}{\varepsilon_i}
\stackrel{\varepsilon_i\to0}
\longrightarrow \frac12 \sum_i p_i^\mu \,{\rm sign}\,(t-t_i),
\qquad\varepsilon_i= \varepsilon \e^{-\varphi_i}.
\label{33}
\ee
An embedding space image of thus smeared stepwise function is a
 {polygon} with vertices
\be
x^\mu_i=\frac 1\K P^\mu_i\,, \qquad 
x^\mu_{i}-x^\mu_{i-1}=\frac 1\K p^\mu_i \,,
\ee 
as is depicted in Fig.~\ref{fi:sme}.
This looks similar to the Wilson-loop/scattering-amplitude duality for 
${\cal N}=4$ SYM advocated by
Alday, Maldacena (2007) \cite{AM07},
Drummond, Korchemsky, Sokatchev (2008) \cite{DKS08},
Brandhuber, Heslop, Travaglini (2008) \cite{BHT08}.

\tit{Invariant regularization and the Liouville field\label{ss:r}}


The invariant regularization used in the previous subsection plays
a very important role for the consistency of scattering amplitudes.
For smeared stepwise $p^\mu(t)$ the Gaussian exponent reads
\bea
&&\hspace*{-24mm}\mbox{}-\pi\int\nolimits_{-\infty}^{+\infty} \d t_1\, 
 \d t_2 \, 
\dot p(t_1) \cdot \dot p(t_2) \, G\left(s(t_1),s(t_2)\right) \non
&=& 
\sum_{k\neq l}\;  p_k \cdot p_l \log |\t_k-\t_l| 
- \pi \sum_j  p_j^2 G(s_j,s_j) .
\eea
The Green function  $G(s_i,s_j)$ has to be regularized
at coinciding arguments, say for 
\mbox{$|s_i-s_j|\la \varepsilon_i,\varepsilon_j$}, which must
be done compatible with the boundary metric for the regularization to be 
invariant. Therefore $G(s_j,s_j)$ has to involve
the Liouville field ${\varphi}$ (${g_{ab}=\e^{\varphi} \delta_{ab}}$)
as pointed out by Polyakov (1981) \cite{Pol81}:
\bea
&&G(s_i,s_j)= -\frac1\pi \ln |s_i-s_j|\qquad \hbox{for } 
|s_i-s_j|\gg \varepsilon_i,\varepsilon_j\,,
\non &&
 G(s_j,s_j) \longrightarrow  G_\varepsilon (s_j,s_j) 
=\frac1\pi \log \frac 1\varepsilon + \frac{1}{2\pi} \varphi(s_j,0)\,.
\label{dos}
\eea

For {critical} open {bosonic string} (in $d=26$) 
Aoyama, Dhar, Namazie (1986) \cite{ADN86} wrote explicitly
\be
A \left( p_1,\ldots, p_M \right) 
= \int \D \varphi(s) \int \prod _{m} \d s_m
\e^{\varphi(s_m,0)/2-\pi \alpha' p_m^2 G(s_m,s_m)}\! 
\prod_{j\neq m} \!|s_j-s_m|^{\alpha' p_j\cdot p_m},
\ee
where the integration over $s_m$ also involves the boundary metric
$\e^{\varphi/2}$. In view of \eq{dos}
the path integration over $\varphi(s,0)$ --- the boundary value 
of the {Liouville field} --- {decouples}, but only on shell, 
\ie for {tachyonic scalar, massless vector, etc.}.
For off-shell amplitudes it has to be properly taken into account
as is described in this talk. 

\tit{Classical (long-string) limit}

At the classical level we are interested in the 
saddle-point approximation to \eq{A[p]}.
The minimal surface spanned by a rectangle with stepwise $p^\mu(t)$ 
displayed in \eq{33} is given by the harmonic function
\be
X^\mu(x,y)=  \frac 1{\pi\K}\sum_i  p_i^\mu  
\arctan \frac{(x-s_i)}{y} \qquad s_i=s(t_i),
\ee
as found in Y.M. (2011) \cite{Mak11a}. It is 
{$T$-dual} to a more familiar (pure imaginary) one
\be
X^\mu(x,y)=  \frac \i{2\pi\K}\sum_i  p_i^\mu  
\ln \left[(x-s_i)^2+y^2\right] \qquad s_i=s(t_i)
\ee
 in coordinate space, known from the early days of string theory \cite{HSV70}.

{Douglas' minimization} with respect to $s_i$'s results (for
$p_i\cdot p_j \gg p_i^2,p_j^2$) in the set of equations
\be
\sum_{j\neq i} \frac{ p_i \cdot p_j}{s_i-s_j}=0.
\label{Dm}
\ee
Only $M-3$ of these are independent 
because of the $PSL(2;\mathbb R)$ 
projective invariance. 
 
For $M=4$ we set $s_1=0$, $s_3=1$,
$s_4=\infty$ in the usual way and obtain
\be
s_2\mbox{}\,_*=\frac{s}{s+t}. 
\ee
Otherwise the projective-invariant ratio is fixed to be 
\be
\left(\frac{s_{21} s_{43}}{s_{31}s_{42}}\right)_*= \frac{s}{s+t}.
\ee
This value is well-known from  the saddle point of the Veneziano amplitude 
at large $-s,-t$.

The polygon bounds the {minimal surface} of the area
\be
\K S_{\rm min} = {\alpha'} s \ln \frac s{s+t}
+{\alpha'} t \ln \frac t{s+t}
\stackrel{-s\gg -t}\to -{\alpha'} t \ln \frac st\,,
\ee
whose exponential reproduces the classical {Regge behavior}:
\be
A(s,t)= \e^{-\K S_{\rm min}} \propto s^{\alpha' t}.
\label{claR}
\ee

\tit{Momentum L\"uscher term}

It is easy to account for semiclassical fluctuations about the minimal surface.
To make a connection with the usual computation of the L\"uscher term for
a $T\times R$ rectangle, we perform the 
 Schwarz--Christoffel
map of the upper half-plane onto the rectangle:
\be
\omega(z)= \sqrt{s_{42}s_{31}}
\int_{s_2}^z \frac{\d x} {\sqrt{(s_4-x)(s_3-x)(x-s_2)(x-s_1)}}
\ee
with
\be
R= 2 K\left(\sqrt{1-\r}\right)\stackrel{\r\to1}\to \pi,\qquad
T=2 K\left(\sqrt{\r}\right)\stackrel{\r\to1}\to  \ln\frac{16}{1-\r}\,,
\ee
where 
\be
\r \equiv \frac{s_{43} s_{21}}{s_{42} s_{31}} =\frac s{s+t}\,,
 \quad \quad s_{ij}=s_i-s_j
\label{defr}
\ee
is the projective-invariant ratio.
Therefore, the ratio
\be
\frac T R = \frac {K\left(\sqrt{\r}\right)}{K\left(\sqrt{1-\r}\right)}
\stackrel{\r\to1}\to \frac 1{24\pi} \ln \frac{16 s}t 
\label{TRat}
\ee
is also projective invariant.

Like for the static potential, each degree of freedom results 
in the momentum-space {L\"uscher term} computed by
Janik (2001) \cite{Jan01}, Y.M. (2011) \cite{Mak11a}
\be
\frac{\pi T}{24 R} = \frac {1}{24} \ln \frac{16 s}t ,
\label{eachdeg}
\ee
where in the Regge regime we have used the asymptote \rf{TRat}.

\tit{Semiclassical Reggeon intercept}

There are $(d-2)$ such sets of degrees of freedom
for bosonic string, so we obtain the linear Regge trajectory 
\be
\alpha(t)=\frac{d-2}{24} +{\alpha'} t.
\label{lineaR}
\ee 
Here ${\alpha'} t$ comes from the classical amplitude \rf{claR},
while the intercept comes from exponentiating 
the momentum {L\"uscher term} on the right-hand 
side of \eq{eachdeg}.

In the effective string theory  description of QCD string,
the parameter 
\be
\ln\frac1{1-\r}=\ln\frac st
\ee 
plays for scattering amplitudes the role of $T$ for the static potential. 
The Regge behavior is like the area law:
\be
A \propto \e^{\alpha(t) \ln (s/t)}\qquad \hbox{is {similar} to}\qquad
W \propto \e^{-T V(R)},
\ee 
while the expansion of $\alpha(t)$ in $(-1/t)$ is like the expansion 
of $V(R)$ in $1/R$.
This is one more manifestation of the Wilson-loop/scattering-amplitude duality.

The semiclassical correction to the 
Regge trajectory of the effective string theory in $d< 26$ 
can be alternatively computed for the upper-half plane parametrization,
repeating that of {{Durhuus, Olesen, Petersen (1984)}} \cite{DOP84} 
for the static potential of the {Polyakov string}. 
Now the same result emerges as
\be
\alpha(0)=1+ \frac{d-26}{24}=\frac{d-2}{24},
\ee
where 1 comes from the boundary Liouville field and $(d-26)/{24}$ comes
from the bulk fluctuations of the effective string.

We thus see the important difference in how the {L\"uscher term} emerges for
the worldsheet and upper-half plane parametrizations: for the former it
is due to quantum fluctuations of $X^\mu (\tau,\sigma)$, while for the latter 
it comes entirely from the classical part of the Liouville field
(= the induced metric).


\tit{Mean-field approximation}

To sum up all orders in $-1/t$, we apply 
the mean-field
method, which works very well for the static potential (for a review see
Y.M. (2012) \cite{Mak12}). 

The (variational) mean-field ansatz with fluctuations included reads
\be
X^\mu(x,y)=  \frac 1{\pi\K}\sum_i  p_i^\mu  
\arctan \frac{(x-s_i)}{y} +X^\mu_{\rm q}(x,y).
\ee
The momenta $p_i$'s and, correspondingly, the  
Mandelstam variables $s$ and $t$ are fixed,
while the ratio $\r$ [defined in \eq{defr}] is a variational parameter:
\be
S_{\rm mf}={\alpha'} s \ln \r 
+ {\alpha'} t \ln (1-\r) +
\frac{(d-2)}{24}\ln(1-\r) \qquad \hbox{valid as $\r\to1$}.
\ee
The first two terms on the right-hand side result from the 
classical quadratic action (in the conformal gauge), while
the third term is the momentum L\"uscher term computed by
Janik (2001) \cite{Jan01}, Y.M. (2011) \cite{Mak11a}.

Minimizing with respect to $\r$, we find
\be
\r_*=1- \frac{{ \alpha'}t+(d-2)/24}{{ \alpha'}s}
\ee
which results in the linear Regge trajectory
\be
\alpha(t)=\frac{(d-2)}{24}+{\alpha'} t ,
\ee
coinciding with the semiclassical one \rf{lineaR}.

The mean field usually works at large $d$, but is expected to be exact for 
bosonic string at any $d$. The arguments are given in Y.M. (2011) \cite{Mak11b}.
Quadratic fluctuations about this mean field are stable for
$\alpha(t) <0$, that is
$-{\alpha'} t > (d-2)/{24}$.
We can go beyond this domain by an analytic continuation.

\section{From bosonic string to QCD}

\tit{Consistent off-shell amplitudes\label{ss:l}}

We have already seen in Subsect.~\ref{ss:r} that off-shell scattering 
amplitudes involve the reparametrization path integral. 
Choosing the Lovelace choice \rf{Lovem} of the discretization of  
the measure and of the Green function at coinciding arguments:
\be
\D^{(N)}_{\rm diff} \t = \prod_{i=1}^N \frac{\d \t_i\,(\t_{i+1}-\t_{i-1})}
{(\t_{i+1}-\t_{i})(\t_{i}-\t_{i-1})} 
\label{measure11}
\ee  
and
\be
G\left(s_j,s_j\right)
\stackrel{{{\rm Lovelace}}}=\frac{1}{\pi}
\ln \frac{(s_{j+1}-s_{j-1})}{(s_{j+1}-s_{j})(s_{j}-s_{j-1})\varepsilon}\,,
\ee
it is possible to integrate over $s_i$'s at the intermediate points,
at which $p^\mu(t)$ has no discontinuities.

As is shown by {Y.M., Olesen (2010)} \cite{MO10b}, this 
results in the scattering amplitude
\be
A( p_1,\ldots, p_M)=
\!\!\!\!\!\!
\int\limits_{\t_{i-1}<\t_i} \prod_{i=1}^M{\d \t_i}\,
\prod_{k\neq l}^M |\t_k-\t_l|^{{\alpha^\prime} p_k\cdot  p_l }
\prod_{j=1}^M
\left(\frac{|\t_j-\t_{j-1}|\,|\t_{j+1}-\t_j|}{|\t_{j+1}-\t_{j-1}|}\right)
^{{\alpha^\prime} p_j^2-1} .
\label{LoveLace}
\ee
The remaining integration over $\t_i$'s (the Koba--Nielsen variables),
at which  $p^\mu(t)$ has discontinuities,
is inherited from the path integral over {reparametrizations}.

Remarkably, the off-shell amplitude \rf{LoveLace} is
a consistent off-shell projective-invariant tree string amplitude
known from Di~Vecchia, Frau, Lerda, Sciuto (1988) \cite{VFLS88}.
It is {invariant} under $PSL(2;\mathbb{R})$ for arbitrary $p_i^2$.
For the tachyonic case, when all $p_i^2=1/\alpha' $, the last
factor on the right-hand side of \eq{LoveLace} equals 1 and 
the amplitude \rf{LoveLace} reproduces the Koba--Nielsen one.

\tit{Application to QCD}

Meson scattering amplitudes in QCD can be extracted from 
{Green's functions} of $M$ {colorless} composite {quark operators} 
of the type
\be
{\bar q}(\x_i) q (\x_i), \quad {\bar q}(\x_i) \gamma_5 q (\x_i),
\quad {\bar q}(\x_i) \gamma_\mu q (\x_i),
\quad {\bar q}(\x_i) \gamma_\mu \gamma_5 q (\x_i),\quad \hbox{etc.}
\ee 
These Green functions are given by the sum
over Wilson loops, passing via the points $\x_i$ ($i=1,\ldots,M$)
at which the operators are inserted:
\be
G\equiv \LA \prod_{i=1}^M \bar{q} (\x_i) q (\x_i) \RA_{\rm conn} 
\stackrel{{\rm large}~\N}= 
\sum_{{\rm paths}\;\ni\, \{\x_1,\ldots,\x_M\equiv \x_0\}}
{J[x(\tau)]} \, W[x(\tau)].
\ee
Here the Wilson loop $W[x(\tau)]$ is in pure Yang--Mills at the large 
number of colors $\N$ (or {quenched}).
This important observation by {Y.M., Migdal (1981)} \cite{MM81}
is a consequence of the large-$\N$ factorization.
The correlators of several Wilson loops are present at finite $\N$.

There exist many ways of representing
the {weight} for the path integration in the worldline formalism. 
I shall use a momentum-space disentangling of gamma-matrices, where
\be
{J[x(\tau)]}=\int \D k(\tau)\;{\rm sp} \;\boldsymbol{P}
\e^{\i \int_0^{\Tau} \d \tau\,
[\dot x (\tau)\cdot k(\tau)-\gamma\cdot k(\tau)]} 
\ee
for {spinor} quarks 
and scalar operators. 
Here $\tau$ is the {proper time} and sp is the trace of 
the path-ordered product of gamma-matrices.
This represen\-ta\-tion is most convenient for dealing with the
momentum loops as we shall shortly see.

Doing the functional Fourier transformation \rf{MigFF},
we obtain, as shown by {Y.M., Olesen (2008)} \cite{MO08},
the following representation for the meson scattering amplitude
\be
A\left( p_1,\ldots,  p_M \right)=
\sum_{\rm paths}
\e^{\i \int_0^\Tau \d \tau\, \dot x (\tau)\cdot p(\tau)}\,
{J[x(\tau)]} \, W[x(\tau)] \,,
\label{66}
\ee
where $p^\mu(\tau)$ is the {piecewise constant}  {momentum loop} \rf{stepW}.

Substituting for  $W[x(\tau)]$ the reparametrization path
integral \rf{W[x]}  and interchanging the integrals 
over $x(\tau)$ ({Gaussian}) and $s(\tau)$, we find 
\begin{eqnarray}
\lefteqn{A\left(p_1,\ldots, p_M \right) \propto
\int\nolimits_0^\infty \d \Tau\, \Tau^{M-1} \e^{-m \Tau} 
\int\nolimits_{-\infty}^{+\infty} \frac{\d \t_{M-1}}{1+\t_{M-1}^2}
\prod_{i=1}^{M\!-\!2}\int\nolimits_{-\infty}^{\t_{i\!+\!1}} \frac{\d \t_i}{1+\t_i^2}}\non 
&& \times \int \D k(t)\;{\rm sp\;} {\boldmath P}
\e^{-\i\Tau  \int \d t\,\gamma\cdot k(t)/(1+t^2)} \,
W\Big[x(t)=\frac1K \left(p(t)+k(t)\right)  \Big].
\end{eqnarray}
It looks like 
the stringy amplitude \rf{A[p]}, 
but with an additional path integral over $k^\mu(\tau)$.

The latter path integral over $k^\mu(\tau)$ remarkably factorizes for 
{small} quark mass $m$%
\footnote{A similar observation was first made by Janik, Peschanski (2002)
 \cite{JP02} (see also Giordano, Peschanski (2011) \cite{GP11}),
using a different representation of the spinor weight.}
and/or {large} $M$ since the integral over $\Tau$  
is {dominated} by {large} $\Tau\sim (M-1)/m$ and typical $k\sim 1/\Tau$.
We thus obtain just the same 
{Lovelace-type} {string amplitude} 
\be
A\left(p_1,\ldots, p_M\right) \propto 
W\Big[x(t)=\frac1\K p(t)  \Big]
\ee
as discussed in the previous subsection.

This result, however, cannot be exact for QCD string since
the reparametrization path integral \rf{W[x]} applies only for large
loops. Perturbative QCD applies instead for small loops.
Nevertheless, large loops dominate the path integral over $x^\mu(\tau)$
in \eq{MigFF} in the Regge kinematical regime, 
as is shown by Y.M., Olesen (2012) \cite{MO12}. This is a pure stringy
phenomenon that large momenta are associated with large distances.
In perturbation theory there is no dimensional parameter like the string
tension and large momenta are associated with small distances.

\tit{Effective $\rho$-trajectory and pQCD prediction}

The above results on the linear Reggeon trajectory in large-$\N$ QCD
can be compared with experiment. We should bear in mind that the 
string disc amplitude is associated with planar graphs and therefore with
the $\rho$-meson type Regge trajectories. It is to be distinguished
from the Pomeron trajectory, which is associated with cylinder graphs.

{The effective $\rho$-meson trajectory, extracted from
the exclusive process $\pi^-p\to \pi^0n$, is reproduced in Fig.~\ref{fi:rho}. 
\begin{figure}
\centerline{\includegraphics[width=7.5cm]{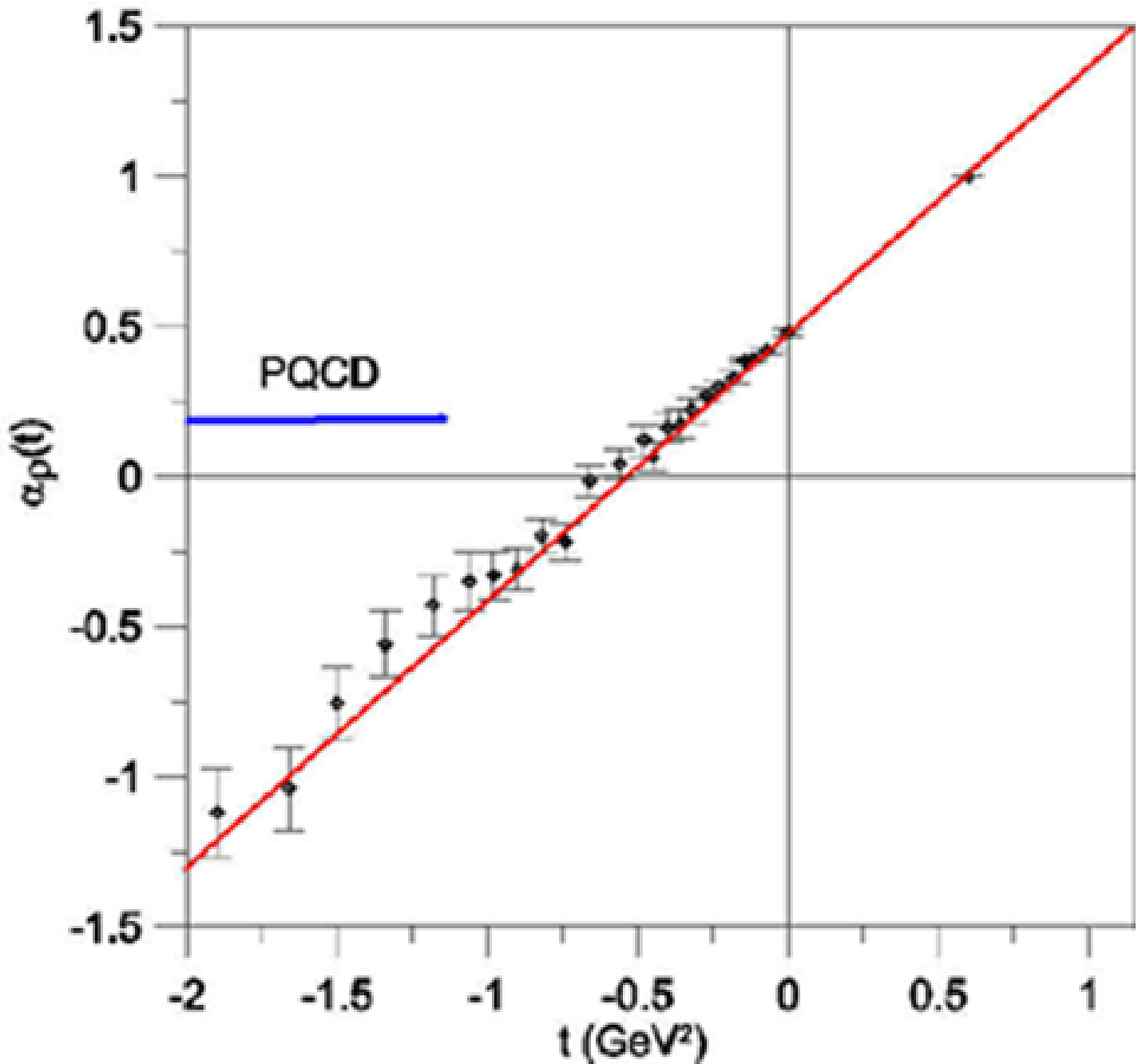} }
\caption{Effective $\rho$ meson trajectory extracted from
the exclusive process $\pi^-p\to \pi^0n$. 
The figure taken from Kaidalov (2006) \cite{Kai06}.}
\label{fi:rho}
\end{figure} 
The figure is taken from Kaidalov (2006) \cite{Kai06}.  
It is seen in the figure that the linear trajectory
of the intercept $\approx 0.5$, known from the spectrum for positive $t$,
continues to negative $t$ down to $t\approx -2~ \hbox{GeV}^2$.
The (almost horizontal) line to the left in Fig.~\ref{fi:rho} corresponds
to perturbative QCD reggeization of  $\bar q q$ by
{Kirschner, Lipatov (1983)} \cite{KL83b}, 
which is expected to work
for $t \la - \hbox{few \ GeV}^2$. As we argue in the next subsection,
such a behavior of the experimental results is because the value
of $s$ was not large enough.

\tit{Separation of pQCD and QCD string}

Let us consider a very simple model of QCD string, when
the Wilson loop equals 1 for  $\K S_{\rm min}<1$ and is
given by \eq{W[x]} for  $\K S_{\rm min}>1$ .
We have thus disregarded gluon interactions and restricted ourselves with
the free contribution for small loops. The motivation is that the
interaction is then small (owing to asymptotic freedom) and the quark
counting rule (the Bjorken scaling) applies for large $-t$, resulting
in the Reggeon trajectory $\alpha(t)=0$. For large loops we substitute
a dual description by the string disk amplitude~\rf{W[x]}.
An analogy with the AdS/CFT correspondence is as if we were substitute
either the Wilson loop in ${\cal N}=4$ super Yang--Mills at small values
of the coupling constant or IIB open superstring in $AdS_5\times S^5$
at large couplings. 
\begin{figure}
\centerline{
\includegraphics[width=11cm]{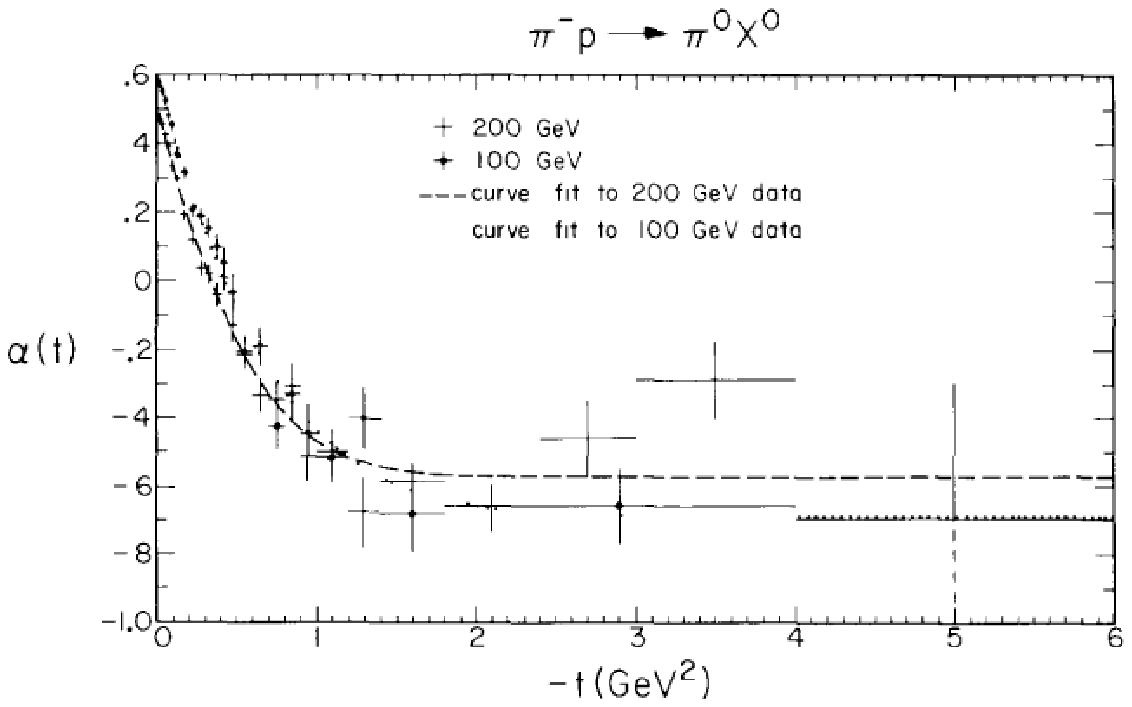}} 
\vspace*{0.7cm}
\centerline{
\hspace*{9mm}\includegraphics[height=6.5cm,width=10cm]{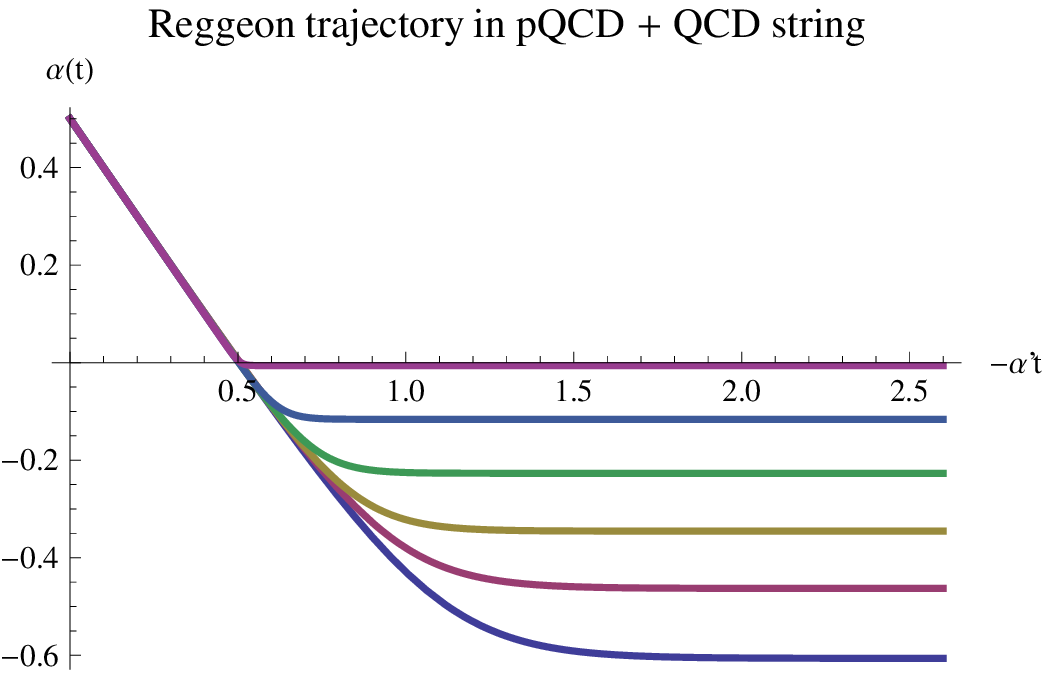} 
}
 \caption{Experimental data by Kennett {\it et al.} (1986) \cite{Ken86} (top)
and their description by pQCD + QCD string (bottom).
The smaller the value of $s$ the lower is the line.}
\label{fi:exrho}
\end{figure} 

We can yet simplify the model, separating large and small loops by
the value of their length rather than the minimal area. This is
legitimate because QCD string is smooth (not crumpled) so the typical
values of the minimal area are proportional to the length squared,
$S_{\rm min}\propto L^2$. We thus substitute
\be
W(C)=\left\{  
\begin{array}{r}
1\quad \hbox{for}~ \sqrt{\K} L<1\,, \\
\hbox{\eq{W[x]}}\quad \hbox{for}~ \sqrt{\K} L>1 \,.
\end{array}
\right.
\label{2fold}
\ee
In practice this means that when the proper time
$\Tau $ is smaller than
$\tau_{\rm max}\sim 1/\K$ we pick up the contribution
from small loops and when it is larger than
$\tau_{\rm max}\sim 1/\K$ we pick up the contribution
from large loops.  Thus $\tau_{\rm max}\sim 1/\K$ plays the role of
an {infrared cutoff} in perturbative QCD instead of the usual 
{transverse mass} $\mu$. 

A nice feature of the ansatz \rf{2fold} is that
the meson scattering amplitude \rf{66} can now be straightforwardly 
computed and equals the sum of the contribution of perturbative
QCD and QCD string with a certain relative coefficient,
which is of most importance at finite $s$.
It is the only parameter to be fixed below by comparing with
experimental data. At infinite $s$ there remains the
contribution from either perturbative QCD or QCD string, depending
on whose exponent is larger at given $t$. However, at finite $s$
both contributions are essential and the lower $s$ is the more
essential the QCD string contribution becomes at 
$-t \approx \hbox{few Gev}^2$.


In Fig.~\ref{fi:exrho} we compare the experimental data
by {Kennett {\it et al.} (1986)} \cite{Ken86} for the
$\rho$-meson trajectory, extracted from
the inclusive process $\pi^-p\to \pi^0 X^0$, 
with the prediction of the model~\rf{2fold}.
The relative coefficient is fixed to fit the data at $s=400~\hbox{GeV}^2$
(the lower line in the bottom figure). The upper the line is
the larger $s$ is. As $s\to\infty$ the two regimes separate. 
The model we have considered quantifies the idea of
Brodsky, Tang, Thorn (1993) \cite{BTT93} about the mixing of
the two regimes in QCD at finite $s$.

\section{Conclusion and Outlook}

The Regge behavior of meson scattering amplitudes 
can be derived for QCD string under practically
the {only assumption} that $\N$ is large. 
Great simplifications occur for small quark mass and/or large 
number of colliding mesons.

It was crucial for the success of calculations that all integrals are
{Gaussian} except for the one over {reparametrizations} which reduces to
the integration over the {Koba--Nielsen variables}.

The mean-field approximation results for QCD string in the linear 
Reggeon trajectory of the intercept  $\alpha(0)=(d-2)/24$.
The actual Reggeon intercept of $\alpha(0)\approx 0.5$ has to be obtained,
most probably, by accounting for 
spontaneous breaking of chiral symmetry.

When $-t \ll s$ becomes large, 
there are {no longer} reasons to expect the contribution of QCD string to 
dominate over {perturbation theory}.
The relative contribution of the two changes with increasing $s$.%
\footnote{It would be most interesting to extend this kind of consideration
to the singlet channel, where the next order of pQCD is
known for the BFKL (Balitsky--Fadin--Kuraev--Lipatov) Pomeron.}

At the end I shall mention some existing extensions of Douglas' 
boundary functional 
and the reparametrization path integral to

\begin{itemize}
\vspace*{-7pt}
\addtolength{\itemsep}{-7pt}

\item closed string (with a possible application to gravity)
by Caputa, Hirano (2012) \cite{CH11},

\item the RNS (Ramond--Neveu--Schwarz) 
superstring by Caputa (2012) \cite{Cap12},

\item the GS (Green--Schwarz) superstring in the
$AdS_5 \times S^5$ background 
(dual to ${\cal N}=4$ super Yang--Mills) by 
Ambj\o rn, Y.M. (2012) \cite{AM11} and Kristjansen, Y.M. (2012) \cite{KM12}.

\end{itemize}

\end{document}